\documentclass[doublecol]{epl2}

\usepackage{graphicx}
\usepackage{amsmath}
\usepackage{amssymb}

\newcommand{\Tk}{T_{\rm K}}
\newcommand{\Geff}{\Gamma_{\rm eff}}

\title{From Weak- to Strong-Coupling Mesoscopic Fermi Liquids}
\shorttitle{From Weak- to Strong-Coupling Mesoscopic Fermi Liquids} 

\author{Dong E. Liu\inst{1}, S\'{e}bastien Burdin\inst{2}, Harold U. Baranger\inst{1}, and Denis Ullmo\inst{3}}
\shortauthor{Dong E. Liu \etal}

\institute{                    
  \inst{1} Department of Physics, Duke University, Box 90305, Durham, North Carolina 27708-0305, USA\\
  \inst{2} Condensed Matter Theory Group, LOMA, UMR 5798, Universit\'{e} de Bordeaux I, 33405 Talence, France\\
  \inst{3} Univ. Paris-Sud, LPTMS UMR 8626, 91405 Orsay Cedex, France
}

\abstract{
We study mesoscopic fluctuations in a system in which there is a continuous connection between two distinct Fermi liquids, asking whether the mesoscopic variation in the two limits is correlated. The particular system studied is an Anderson impurity coupled to a finite mesoscopic reservoir described by random matrix theory, a structure which can be realized using quantum dots. We use the slave boson mean field approach to connect the levels of the uncoupled system to those of the strong coupling Nozi\`eres Fermi liquid. We find strong but not complete correlation between the mesoscopic properties in the two limits and several universal features.
}

\begin{document}
\maketitle

\section{Introduction}
The Fermi liquid is a ubiquitous state of electronic matter \cite{Landau57,PinesNozieres65,KolnKarlsruheRMP07}. Indeed, it is so common that systems can have several different Fermi liquid phases in different parameter regimes (controlled by different fixed points), leading to cross-overs between Fermi liquids with different characteristics. 
Examples of such cross-overs include, for instance, the half-filled Landau level (high-temperature to low-temperature connection) \cite{HalperinLeeReadPRB93}, heavy fermion materials  \cite{HFrecentoverview, KolnKarlsruheRMP07, HewsonBook, BookFulde06}, as well as the simple spin $1/2$ Kondo problem which will be our main concern in this paper. 
In the bulk, clean case, the evolution of the quasi-particles in such a cross-over is straight forward: both sets of quasi-particles are labeled by $\mathbf{k}$ because of translational invariance and so are in one-to-one correspondence. However, in the absence of translational invariance---such as in a disordered or mesoscopic setting---interference affects the two sets of quasi-particles differently. In such a situation, it is interesting to ask how the quasi-particles in one Fermi liquid are related to those in the other. 

The Kondo problem provides a particularly clear example: at weak
coupling (high temperature) the electrons in the Fermi sea are nearly
non-interacting while the strong coupling (low temperature) behavior
is described by Nozi\`ere's Fermi liquid theory
\cite{Nozieres74,HewsonBook}. The connection between high and low
temperature is provided, e.g., by Wilson's renormalization group
calculation \cite{WilsonRMP75}, yielding a smooth cross-over.

We now break translational invariance by supposing that the size of the Fermi sea is finite; it could consist of, for instance, a large quantum dot or metallic nano-particle. The density of states in the electron sea will typically have low energy structure and features, in contrast to the intensively studied flat band case. The finite size effects introduce two additional energy scales: (i)~a finite mean level spacing, leading to what is called the ``Kondo box'' problem \cite{Thimm99,Simon02,Cornaglia02a}, and (ii)~the Thouless energy  $E_{\rm Th} = \hbar/\tau_{c}$ where $\tau_c$ is the typical time to travel across the finite reservoir. When probed with an energy resolution smaller than $E_{\rm Th}$, both the spectrum and the wave-functions of the electron sea display mesoscopic fluctuations, which affect the Kondo physics \cite{KaulEPL05, Yoo05, UllmoRPP08}. Disorder in the electron sea causes similar effects \cite{Kettemann04,Kettemann06,Kettemann07,Zhuravlev08}.

\begin{figure}[t]
\centering
\includegraphics[width=3.1in,clip]{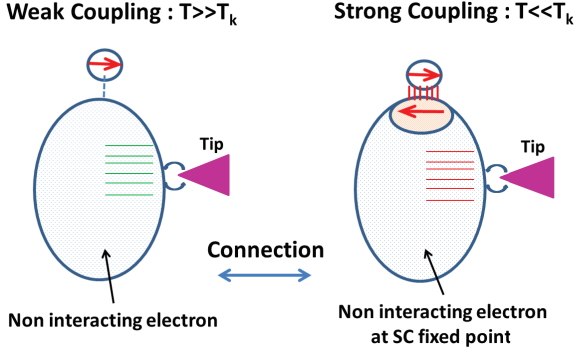}
\caption{(Color online) Schematic illustration of small-large quantum dot system. Left panel : weak coupling limit $T \gg\Tk$. Right panel : strong coupling limit $T \ll \Tk$. The energy levels and wave functions probed by the tip change from one Fermi liquid regime to the other.}
\label{fig:demo}
\end{figure}

Consider the system shown in Fig.~\ref{fig:demo}: a small Kondo dot coupled to a large ``reservoir dot'' probed weakly by tunneling from a tip. In the high temperature regime, the small dot is weakly coupled to the large dot which is essentially non-interacting. Mesoscopic fluctuations of the density of states translate into mesoscopic fluctuations of the Kondo temperature. Once this translation is taken into account, the high-temperature physics remains essentially the same as in the flat band case \cite{KaulEPL05, Yoo05, UllmoRPP08}; in particular, physical properties can be written as the same universal function of the ratio $T/\Tk$ as in the bulk flat-band case, as long as $\Tk$ is understood as a realization dependent parameter \cite{KaulEPL05, Yoo05, UllmoRPP08}. 
In contrast, the consequences of mesoscopic fluctuations for low temperature Kondo physics ($T \ll \Tk$, strong coupling limit) remain largely unexplored. A few things are nevertheless known: for instance,
the very low temperature regime should be described by a Nozi\`eres-Landau Fermi liquid, as in the original Kondo problem. Indeed, the physical reasoning behind the emergence of Fermi liquid behavior at low temperature, namely that for energies much lower than $\Tk$ the impurity spin has to be completely screened, applies as well in the mesoscopic case as long as $T<\Delta \ll \Tk$ \cite{Affleck01, Simon03, Simon06,KaulPRL06,Kaul09,Pereira08}. In this case, the system consists of the Kondo singlet plus non-interacting electrons with a $\pi/2$ phase shift as shown in the right panel of Fig.~\ref{fig:demo}.

Measurements of the conductance through the large dot or the ac response to the tip reveal the mesoscopic fluctuations of the energy levels and wavefunctions \cite{Kouwenhoven97, BirdQdotsBook03}. Thus, such experiments can probe the connection between the quasi-particles in the two Fermi liquid regimes, as well as the properties of the intermediate strongly correlated Kondo cloud \cite{Simon06,KaulPRL06,Kaul09,Pereira08}. In this paper, we study this connection explicitly, using slave boson mean field (SBMF) theory \cite{ Lacroix79, Coleman83, Read84, KolnKarlsruheRMP07, HewsonBook, BookFulde06, BurdinRev2009, Bedrich10} to treat the interactions and random matrix theory (RMT) \cite{Bohigas91} to model the mesoscopic fluctuations. We find that the correlation between the properties of the two sets of quasi-particles is substantial but not complete. 

\section{Model}
The system pictured in Fig.~\ref{fig:demo} can be described by the Hamiltonian 
$H=H_{\rm bath}+H_{\rm imp}$
where $H_{\rm bath}$ describes the mesoscopic electronic bath and $H_{\rm imp}$ 
describes the local ``magnetic impurity''---small quantum dot, nanoparticle, or magnetic ion---and its interaction with the bath. $H_{\rm bath}$ is the non-interacting Hamiltonian  
$H_{\rm bath}\equiv 
\sum_{i, \sigma}(\epsilon_{i}-\mu)c_{i\sigma}^{\dagger}c_{i\sigma}$,
where $i=1,\cdots ,N$ labels the levels, 
$\sigma=\uparrow,\downarrow$ is the spin component, and  $\mu$ is the
chemical potential. $H_{\rm imp}$ is
\begin{equation}
\label{HimpAndersonboxU1}
H_{\rm imp}=
V_0\sum_{\sigma}[c_{0\sigma}^{\dagger}d_{\sigma}+
d_{\sigma}^{\dagger}c_{0\sigma}]
+E_{d}\sum_{\sigma}d_{\sigma}^{\dagger}d_{\sigma}
\end{equation}
where the $d_\sigma$ operators refer to the impurity site with energy $E_d$ and the position of the impurity is taken to be $\mathbf{r}=0$. We take the local Coulomb interaction between $d$-electrons to be $U=\infty$; thus, 
states with two $d$-electrons must be projected out. Finally,
the local electronic operator $c_{0\sigma}$ is related to the bath eigenstate operators $c_{i\sigma}$ through 
$c_{0\sigma}=\sum_{i=1}^{N}\phi_{i}^{*}(0)c_{i\sigma}$
where $\phi_i({\bf r}) = \langle{\bf r}| i \rangle $ are the one-body wave functions of $H_{\rm bath}$ with the local normalization relation $\sum_{i}\vert \phi_{i}(0)\vert^{2}=1$.

To study the mesoscopic fluctuations, we assume that the classical
dynamics within the large dot is chaotic, and thus that the energy
levels $\epsilon_{i}$ and the wave functions at the impurity site
$\phi_{i}(0)$ are described by random matrix theory (RMT)
\cite{Bohigas91,UllmoRPP08,Kouwenhoven97}, specifically, by the
Gaussian orthogonal ensemble (GOE) for time reversal symmetric systems
and the Gaussian unitary ensemble (GUE) for systems in which time
reversal is broken \cite{MehtaBook,Bohigas91}.

Applying the SBMF approximation \cite{ Lacroix79, Coleman83, Read84,
  KolnKarlsruheRMP07, HewsonBook, BookFulde06, BurdinRev2009,
  Bedrich10}, we introduce auxiliary boson $b^{\dagger}$ and fermion
$f_{\sigma}^{\dagger}$ operators, such that $d_{\sigma} =
b^{\dagger}f_{\sigma}$, with the constraint $b^\dagger b +
\sum_{\sigma}f^{\dagger}_{\sigma}f_{\sigma}=1$.  Since the Hamiltonian
is invariant with respect to a $U(1)$ gauge transformation, the
bosonic field can be treated as a real number : $b, b^{\dagger}
\mapsto \eta$. The constraint condition is satisfied by introducing a
static Lagrange multiplier, $\xi$.  One thus obtains the SBMF
effective Hamiltonian
\begin{eqnarray}
\label{MFHamiltonianAndersonbox1} 
H_{\rm MF} &=& \sum_{\sigma}\Big\{ 
\sum_{i=1}^{N}(\epsilon_{i}-\mu)c_{i\sigma}^{\dagger}c_{i\sigma} 
+(E_{d}-\xi)f_{\sigma}^{\dagger}f_{\sigma}  \nonumber\\
&&+\eta V_0 (c_{0\sigma}^{\dagger}f_{\sigma} 
 + f_{\sigma}^{\dagger}c_{0\sigma}) \Big\}
+ \xi(1-\eta^{2}) \;. 
\end{eqnarray}
The mean field parameters $\eta$ and $\xi$ are obtained by minimizing
the free energy of the system, taking $\mu=0$. Using the equations of
motion from the mean-field
Hamiltonian~Eq.\,(\ref{MFHamiltonianAndersonbox1}), we obtain, after
some algebra, the imaginary time Green function
\begin{equation}
 \label{AndersonboxGff1}
 G_{\rm ff}(i\omega_{n}) =
\Big[ i\omega_{n}+\xi -E_{d}-\eta^{2}V_0^{2}
\sum_{i=1}^{N}\frac{\vert \phi_i(0)\vert^{2}}{i\omega_{n}+\mu-\epsilon_{i}}
\Big]^{-1}
\end{equation}
from which all the properties of the system can be derived.

The eigenvalues $\lambda_\kappa$ and eigenstates $|\psi_\kappa\rangle$
($\kappa = 0,1,\cdots,N$) of the mean field Hamiltonian
Eq.~(\ref{MFHamiltonianAndersonbox1}) correspond to
{the quasi-particles of}
the strong coupling limit.  
{Because the low temperature/energy regime of the system is a Fermi liquid, 
  the mean field approach provides
  a good description of the low energy properties of the strong
  coupling limit, but it is not expected to be accurate at higher energies.  As a
  consequence, it is mainly the range $|\lambda_\kappa - \mu|
  \stackrel{\sim}{<} T_K$ which is physically relevant in terms of
  Kondo physics.  We shall therefore in the following
  concentrate on this energy range.}  
Since the
tunneling strength at energy $E$ between an external tip and the large
quantum dot (see Fig.~\ref{fig:demo}) depends on the line-up of the
levels and the wavefunction intensity, both $\lambda_\kappa$ and
$|\psi_\kappa(\bf r)|^2$ are measurable in experiments.

We now study the relation between the
$\{\lambda_\kappa, \! |\psi_\kappa\rangle\}$ and the
$\{\epsilon_i, \! |\phi_i\rangle\}$.  Expressing the Green function of
$H_{\rm MF}$ as
\begin{equation} \label{eq:Green}
\hat G(\lambda-\mu) = [\lambda-\mu-H_{\rm MF}]^{-1}
=
\sum_0^N \frac{|\psi_\kappa\rangle \langle
  \psi_\kappa|}{\lambda - \lambda_\kappa} \; ,
\end{equation}
one sees that $(\lambda_\kappa-\mu)$  are the poles  of the Green function 
$G_{\rm ff}(z) =\langle f | \hat G(z)
|f\rangle$. Eq.~(\ref{AndersonboxGff1}) then immediately implies that
the 
$\lambda_\kappa$ are solutions of the equations
\begin{equation} \label{eq:lambdas}
\frac{\Delta}{\pi}  \sum_{i=1}^{N}\frac{|\phi_i(0)|^2}{\lambda-\epsilon_{i}} 
=
\frac{\lambda - \mathcal{E}_0(\xi)}{\Geff} 
\; ,
\end{equation}
where $\mathcal{E}_0(\xi)  \equiv  E_d + \mu - \xi$ 
(interpreted as the position of the Kondo resonance if the system is in the Kondo regime) and
$\Geff \equiv   \pi \rho_0 \eta^2 V_0^2$ (interpreted as the width of the Kondo resonance, which gives the scale of the Kondo temperature). 
$\rho_0=1/\Delta$ is the mean density of states. Note that 
Eq.~(\ref{eq:lambdas}) implies that there is one and
only one $\lambda_\kappa$ in each interval
$[\epsilon_i,\epsilon_{i+1}]$.

The probability of overlap between the eigenstate $\kappa$ and the impurity state $| f\rangle$, $u_\kappa \equiv |\langle f |\psi_\kappa\rangle |^2$, is a key ingredient in how the wave function amplitude at $\bf r$ is affected by the Kondo singlet. Since the $u_\kappa$ are the residues of $G_{\rm ff}(z)$, Eq.~(\ref{AndersonboxGff1}) implies
\begin{equation} 
\label{eq:uk}
u_\kappa = \left[ 1 +
\displaystyle
  \frac{\Gamma_{\rm eff}}{\pi} \sum_{i=1}^N \frac{|\phi_i(0)|^2
    \Delta}{(\lambda_\kappa - \epsilon_i)^2} \right]^{-1} \; . 
\end{equation}

For $|\lambda -\mathcal{E}_0(\xi)| \gg \Gamma_{\rm eff}$, one
contribution dominates the sum on the left hand side of Eq.~(\ref{eq:lambdas})---namely, the closest $\epsilon_i$ to $\lambda_{\kappa}$, call it $i(\kappa)$---in which case $\lambda_\kappa = \epsilon_{i(\kappa)} + \delta_\kappa \Delta$ with
$\delta_\kappa \simeq \Gamma_{\rm eff} |\phi_{i(\kappa)}(0)|^2 /
[\pi (\lambda_\kappa - \mathcal{E}_0 )] \ll 1$.
As expected, the two spectra nearly coincide. Similarly, the participation of the wavefunctions in the singlet state is small: from Eq.~(\ref{eq:uk})
\begin{equation} \label{eq:uk_out}
u_\kappa \simeq \frac{\Gamma_{\rm eff}}{\pi} \frac{|\phi_{i(\kappa)}(0)|^2 \Delta}{(\lambda_\kappa
  - \mathcal{E}_0)^2} \ll \frac{\Delta}{\Gamma_{\rm eff}} \; .  
\end{equation}

In contrast, for $|\lambda -\mathcal{E}_0(\xi)| \ll \Gamma_{\rm eff}$, 
the right hand side of Eq.~(\ref{eq:lambdas}) can be neglected. 
The typical distance between a $\lambda_\kappa$ and the closest 
$\epsilon_i$ is then of order $\Delta$, and 
$u_\kappa \sim \Delta/\Gamma_{\rm eff}$.
In the limit $\Tk \sim \Gamma_{\rm eff} \gg \Delta$, only the wave function amplitudes for energy levels within the Kondo resonance, 
$|\lambda_\kappa -\mathcal{E}_0(\xi)| \ll \Gamma_{\rm eff}$, will be significantly affected. 

\section{Energy Spectral Correlation}
\begin{figure}[t]
\centering
\includegraphics[width=3.4in,clip]{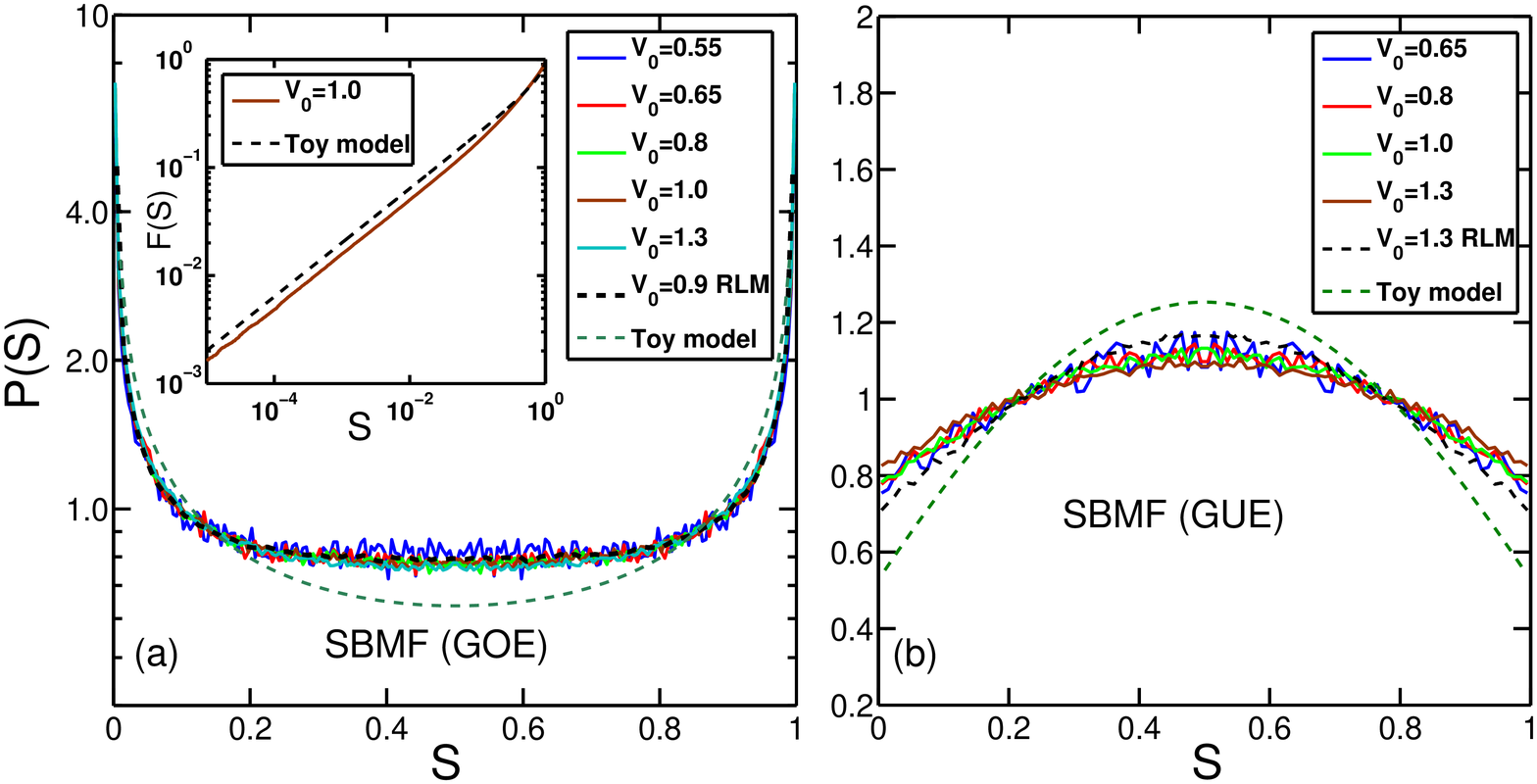}
\vspace*{-0.2in}
\caption{(Color online) The distribution of $S$, including both
  $|\lambda_\kappa-\epsilon_i|/|\epsilon_{i+1}-\epsilon_i|$ and
  $|\lambda_\kappa-\epsilon_{i+1}|/|\epsilon_{i+1}-\epsilon_i|$, from
  the SBMF treatment of the infinite-$U$ Anderson model; (a) GOE, (b)
  GUE.  The dashed lines are the results of the toy model. Inset (a):
  the cumulative distribution {$F(S) \equiv
      \int_0^S p(x) d x $} of the $V{_0}=1.0$ GOE data
    compared to the toy model. Note the presence of the square-root
    singularity.  Parameters: full band width $D=3$, impurity energy
    level $E_d=-0.7$, 500 energy levels within the band, and 5000
    realizations used.
}
\label{fig:LevelCor}
\end{figure}

To characterize the relation between the weak and strong coupling
levels, $\{\epsilon_i\}$ and $\{\lambda_\kappa\}$ respectively, we consider the distribution of the normalized level shift defined by 
\begin{equation}
   S \in \left\{ \frac{|\lambda_\kappa-\epsilon_i|}{|\epsilon_{i+1}-\epsilon_i|},
   \frac{|\lambda_\kappa-\epsilon_{i+1}|}{|\epsilon_{i+1}-\epsilon_i|} \right\} 
\end{equation}
where $\epsilon_i$ and $\epsilon_{i+1}$ are the two levels which sandwich 
$\lambda_\kappa$. The range of $S$ is from $0$ to $1$. The probability distribution $P(S)$ obtained numerically using the SBMF approximation by sampling a large number of realizations is shown in
Fig.~\ref{fig:LevelCor} for several cases. Only the levels that are within the Kondo resonance are included; that is, levels satisfying  
$|\lambda_\kappa - \mathcal{E}_0 | < \Geff/2$. 
Note in particular two features of the numerical results: (i) the strong coupling levels are more concentrated near the 
original levels in the case of the GOE while they are pushed away from
the original levels in the GUE, and (ii) the distribution found is completely independent of $V{_0}$. 

An explanation for both of these features can be found from a simple
analytic approximation to the distribution $P(S)$. Well within the
resonance, $|\lambda_\kappa - \mathcal{E}_0| \ll \Gamma_{\rm eff}$,
the r.h.s.\ of Eq.~(\ref{eq:lambdas}) can be set equal to zero, thus
leading to the simplification $ \sum_{i=1}^N
|\phi_i(0)|^2/(\lambda_\kappa - \epsilon_i) \approx 0$.  Focusing on
the level $\lambda_\kappa$ located between $\epsilon_i$ and
$\epsilon_{i+1}$, we consider a toy model in which the influence of
all but these closest $\epsilon$'s is neglected, yielding the much
simpler equation for $\lambda_\kappa$
\begin{equation} 
\label{eq:Connection_toy_bis}
   \frac{|\phi_i|^2}{\lambda_\kappa - \epsilon_i}
+  \frac{|\phi_{i+1}|^2}{\lambda_\kappa - \epsilon_{i+1}}  = 0 \;.
\end{equation}
In RMT, the wavefunction amplitudes $|\phi_i|^2$ and $|\phi_{i+1}|^2$
are uncorrelated and distributed according to the Porter-Thomas
distribution \cite{MehtaBook,Bohigas91}. Notice that all energy scales
($V{_0}$, $\Delta$, etc.) have disappeared from the problem
except for $\delta \epsilon \equiv \epsilon_{i+1} - \epsilon_i$.  The
resulting distribution of $\lambda_\kappa$ is therefore
\emph{universal}, depending only on the symmetry under time
reversal. Hence the empirical observation in Fig.~\ref{fig:LevelCor}
that the curves are independent of $V{_0}$.

Integration over the Porter-Thomas distributions gives
\begin{eqnarray}
P(\lambda_\kappa) &  = &  \frac{1}{\pi} 
\frac{1}{\sqrt{(\epsilon_{i+1} - \lambda_\kappa)(\lambda_\kappa
    - \epsilon_i)}} \qquad \mbox{GOE} \label{eq:PtoyGOE}\\
P(\lambda_\kappa) &  = &  \frac{1}{\delta\epsilon} 
\qquad \qquad \qquad \mbox{GUE} \; . \label{eq:PtoyGUE}
\end{eqnarray}
Breaking time-reversal symmetry thus affects drastically the
correlation between the low temperature levels $\lambda_\kappa$ and
the neighboring high temperature ones $\epsilon_i$ and
$\epsilon_{i+1}$. Time-reversal 
symmetric systems show clustering, with a square root singularity, of
the $\lambda_\kappa$'s close to the $\epsilon_i$'s, while for systems
without time-reversal symmetry the distribution is uniform between 
$\epsilon_i$ and $\epsilon_{i+1 }$. The GUE result can be improved by taking into account the other levels on average; this yields the expression plotted in Fig.~\ref{fig:LevelCor}(b), but as it is lengthy we do not specify it here. Note that this improved toy model does give the bunching of levels in the middle of the interval seen in the numerics.

The difference between $P(S)$ in the two ensembles comes from the very different wave function distribution: the GOE Porter-Thomas distribution has a square root singularity at $|\phi_i(0)|^2=0$ while it is finite for the GUE. The high probability of small wave function amplitudes in the GOE leads to the clustering of strong coupling levels around the original ones. To explore this in the SBMF numerical results,  we plot the cumulative distribution function on a log-log scale in the inset in Fig.~\ref{fig:LevelCor}; the resulting straight line parallel to the toy model result shows that, indeed, the square root singularity is present.

\section{Wave Function Correlations}

A key quantity in quantum dot physics is the magnitude of
the wave function of a level at a point in the dot that is coupled to
an external lead (see Fig.~\ref{fig:demo}). This quantity is directly related to the conductance into the dot when the chemical potential in the lead is close to the energy of the level. We assume that the probing lead is very weakly coupled, so
that the relevant quantity is the wave function in the absence of leads.
To see how the tunneling to an outside lead at $\mathbf{r}$ is affected by the coupling to the impurity, we study the correlation between the strong coupling wave-function intensity $|\psi_{\kappa(i)}(\mathbf{r})|^2$ and its weak coupling counterpart $|\phi_i(\mathbf{r})|^2$, with $\kappa(i) \equiv
i$ for $\lambda_\kappa < \mathcal{E}_0$ and $\equiv (i+1)$ for $\lambda_\kappa > \mathcal{E}_0$. Specifically, we consider the correlator
\begin{equation}
\label{eq:wavefunCorr}
\mathcal{C}_{i,\kappa(i)} = \frac{\overline{|\phi_i(\mathbf{r})|^2
    |\psi_{\kappa(i)}(\mathbf{r})|^2}- 
\overline{|\phi_i(\mathbf{r})|^2}\cdot\overline{|\psi_{\kappa(i)}(\mathbf{r})|^2}}
   {\sigma(|\phi_i(\mathbf{r})|^2)\sigma(|\psi_{\kappa(i)}(\mathbf{r})|^2)}
   \; .
\end{equation}
The average $\overline{(\cdot)}$ here is over all
realizations, for arbitrary fixed $\mathbf{r} \neq 0$, {and
  $\sigma(\cdot)$ is the square root of the variance of the
  corresponding quantity.}

\begin{figure}[t]
\centering
\vspace*{0.2in}
\includegraphics[width=3.4in,clip]{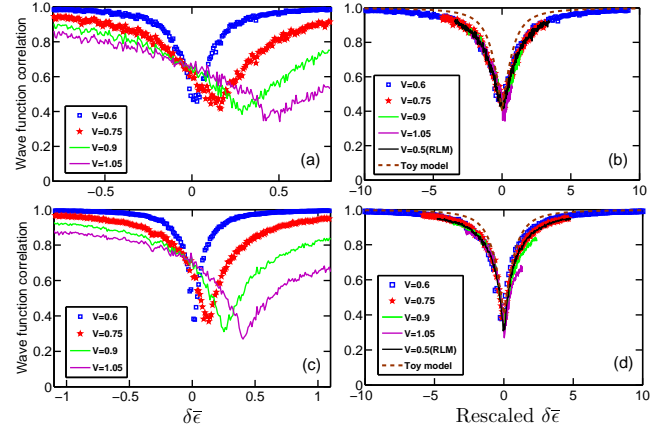}
\vspace*{-0.2in}
\caption{(Color online) Wave function correlation,
  $\mathcal{C}_{i,\kappa(i)}$, for the SBFM approach to the
  infinite-$U$ Anderson model.  
  (a) GOE and (c) GUE, as a function of the average
  distance from the middle of the band. (b) GOE and (d) GUE,
  as a function of the \emph{rescaled} average distance 
  $[i\Delta \!- \!D/2 - (\mathcal{E}_0(\xi)\!
  -\!\mu)]/\Gamma_{\rm eff}$. Dashed line: analytic approximation,
  Eq.~(\ref{eq:Cikappa}).  
  Parameters: full band width $D=3$, impurity energy level $E_d=-0.7$, 
  500 energy levels within the band, and 2000 realizations. 
  }
\label{fig:SBMFT_WF}
\end{figure}

Results for $\mathcal{C}_{i,\kappa(i)}$ from the SBMF approach to the infinite-$U$ Anderson model are shown in Fig.~\ref{fig:SBMFT_WF}. Two ways of showing the dependence on the argument $i$ are used: in the left hand panels, the $x$-axis is simply the (average) energy from the middle of the band, namely $\delta \overline\epsilon_i \equiv i \Delta- D/2$, while in the right hand panels, this energy is scaled so that the $x$-axis is the energy from the center of the Kondo resonance in units of the Kondo temperature (see caption for exact expression). Because the infinite-$U$ Anderson model is inherently not particle-hole symmetric, the location of the Kondo resonance is not at zero but rather increases as $V$ increases so that the average occupation of the impurity level is less than one. 

The scaled curves have a very natural interpretation. First, those states which do not participate in the Kondo singlet state at low temperature, 
$|\delta\overline\epsilon -(E_d-\xi)| \gg \Gamma_{\rm eff}$, are essentially unchanged, $\mathcal{C}_{i,\kappa(i)} \sim 1$. In contrast, those states with energies within the Kondo resonance are substantially changed by interaction with the impurity. The universality of the low energy Kondo physics is nicely demonstrated by the collapse of all the numerical curves for different coupling strengths onto universal curves, one for the GOE and one for the GUE. 

The most interesting feature in Fig.~\ref{fig:SBMFT_WF} is that the correlation does \emph{not} go to zero at the center of the Kondo resonance, even for strong bare coupling. Clearly, the wave functions of the weak coupling and strong coupling Fermi liquid states are similar to each other in that the interference pattern in the original wave function is not completely wiped out by the formation of the Kondo singlet state. This residual correlation should be observable as a correlation in the conductance probed by an external tip.

Expressing the wave function probability $|\psi_{\kappa}(\mathbf{r})|^2$ as the residue of $\hat{G}(\mathbf{r},\mathbf{r})$, and using that in the semiclassical limit the magnitude of the unperturbed wave-functions $\phi_i (\mathbf{r})$ are uncorrelated at different points and with the energy levels, one can show \cite{us-future} that in the limit $\Geff \gg \Delta$, the correlator is given by
\begin{equation}
\mathcal{C}_{i,\kappa} = \overline{\Omega^\kappa_{ii}}
= \overline{u_\kappa \cdot \frac{|v_i|^2}{\lambda_\kappa - \epsilon_i}}
\end{equation}
where $v_i=\eta V_0 \phi_i(0)$. An approximation to $\mathcal{C}_{i,\kappa(i)}$ can be obtained by taking the energy levels to be evenly spaced and replacing the wave function intensities by the average value; this yields
\begin{equation}
\left(\Omega^\kappa_{ii}\right)^{\rm bulk} \equiv
\frac{1}{\displaystyle \delta^{2}_\kappa \sum_i \frac{1}{(i+\delta_\kappa)^2}} \; 
\end{equation}
where $\delta_\kappa \equiv (\lambda_{\kappa(i)} - \epsilon_i)/\Delta$. 
Within the same approximations, Eq.~(\ref{eq:lambdas}) then implies
\begin{equation} \label{app:deltak}
\frac{\lambda_\kappa - \mathcal{\bar E}_0}{\bar \Gamma_{\rm eff}} 
= \frac{1}{\pi}  \sum_{j}\frac{1}{\delta_\kappa-j} = {\rm cotan}(\pi \delta_\kappa) \;. 
\end{equation}
By defining $\bar \lambda_{\kappa} \equiv \lambda_{\kappa} - \mathcal{\bar E}_0$, we obtain for the correlator
\begin{equation} \label{eq:Cikappa}
\mathcal{C}_{i,\kappa(i)} \simeq \frac{1}{\left[{\rm
      cotan}^{-1}\left({\bar \lambda_{\kappa}}/{\bar \Gamma_{\rm eff}} \right)
  \right]^2 \left(1 + \left( {\bar \lambda_{\kappa}}/{\bar \Gamma_{\rm eff}}
    \right)^2 \right)}
  \;
\end{equation}
which, as anticipated, depends only on the ratio $( {\bar \lambda_{\kappa}}/{\bar \Gamma_{\rm eff}})$.  
The curve resulting from this expression is shown in Fig.~\ref{fig:SBMFT_WF} (b) and (d); it yields the value $\mathcal{C}_{i,\kappa(i)} \sim 4/\pi^2$ at the minimum, independent of all 
parameters \cite{us-future}. The value found numerically is slightly smaller but in reasonable agreement.

\section{Conclusion}
We have presented the first study of mesoscopic fluctuations in two distinct but continuously connected 
Fermi liquids by using the slave boson mean field approximation to calculate the strong-coupling levels. 
In the specific case that we study---a small quantum dot coupled to a large reservoir quantum dot with chaotic dynamics---the 
fluctuations of single particle properties in the two limits are highly correlated, universal, and very 
sensitive to time reversal symmetry. Indeed, each strong coupling level must lie between two of the original 
levels (the spectra are interleaved), and, in the GOE case but not the GUE, the levels within the Kondo 
resonance are clustered about the weak coupling ones. Similarly, while the wave function correlation dips 
within the Kondo resonance, it remains substantial, showing that while the corresponding wave function is 
strongly affected by the Kondo screening it retains a surprisingly substantial overlap with the original 
wave function. We expect a similar strong correlation between the properties of continuously connected 
distinct Fermi liquids in other systems. An interesting extension would be to study such correlations when 
a quantum phase transition intervenes as in, e.g., the two impurity Kondo model.

\acknowledgments
The work at Duke was supported by U.S.\,DOE, Office of Basic Energy Sciences, Division of Materials Sciences and Engineering under award \#DE-SC0005237 (DEL and HUB).
SB acknowledges support from the French ANR programs SINUS and IsoTop.
\bibliographystyle{eplbib}

\bibliography{kondo,rmt,nano}

\begin{thebibliography}{10}
\expandafter\ifx\csname url\endcsname\relax\def\url#1{\texttt{#1}}\fi

\bibitem{Landau57}
\Name{Landau L.~D.} \REVIEW{JETP }{3}{1957}{920}.

\bibitem{PinesNozieres65}
\Name{Pines D. \and Nozieres P.} \Book{Theory of Quantum Liquids} (W.A.
  Benjamin, New York) 1965.

\bibitem{KolnKarlsruheRMP07}
\Name{von Lohneysen H., Rosch A., Vojta M. \and Wolfle P.} \REVIEW{Rev. Mod.
  Phys. }{79}{2007}{1015}.

\bibitem{HalperinLeeReadPRB93}
\Name{Halperin B.~I., Lee P.~A. \and Read N.} \REVIEW{Phys. Rev. B
  }{47}{1993}{7312}.

\bibitem{HFrecentoverview}
\Name{Flouquet J., Aoki D., Bourdarot F., Hardy F., Hassinger E., Knebel G.,
  Matsuda T., Meingast C., Paulsen C. \and Taufour V.} \Book{Trends in heavy
  fermion matter} in \Book{International Conference On Strongly Correlated
  Electron Systems (SCES 2010)}, edited by \Name{Ronning F. \and Batista C.}
  (Journal of Physics Conference Series, Volume 273) 2011.

\bibitem{HewsonBook}
\Name{Hewson A.~C.} \Book{The Kondo Problem to Heavy Fermions} (Cambridge
  University Press, Cambridge) 1993.

\bibitem{BookFulde06}
\Name{Fulde P., Thalmeier P. \and Zwicknagl G.} \Book{Strongly correlated
  electrons} in \Book{Solid State Physics} Vol.~60 (Elsevier, New York) 2006
  pp. 1--180.

\bibitem{Nozieres74}
\Name{Nozi\`eres P.} \REVIEW{J. Low Temp. Phys. }{17}{1974}{31}.

\bibitem{WilsonRMP75}
\Name{Wilson K.~G.} \REVIEW{Rev. Mod. Phys. }{47}{1975}{773}.

\bibitem{Thimm99}
\Name{Thimm W.~B., Kroha J. \and von Delft J.} \REVIEW{Phys. Rev. Lett.
  }{82}{1999}{2143}.

\bibitem{Simon02}
\Name{Simon P. \and Affleck I.} \REVIEW{Phys. Rev. Lett. }{89}{2002}{206602}.

\bibitem{Cornaglia02a}
\Name{Cornaglia P.~S. \and Balseiro C.~A.} \REVIEW{Phys. Rev. B
  }{66}{2002}{115303}.

\bibitem{KaulEPL05}
\Name{Kaul R.~K., Ullmo D., Chandrasekharan S. \and Baranger H.~U.}
  \REVIEW{Europhys. Lett. }{71}{2005}{973}.

\bibitem{Yoo05}
\Name{Yoo J., Chandrasekharan S., Kaul R.~K., Ullmo D. \and Baranger H.~U.}
  \REVIEW{Phys. Rev. B }{71}{2005}{201309(R)}.

\bibitem{UllmoRPP08}
\Name{Ullmo D.} \REVIEW{Rep. Prog. Phys. }{71}{2008}{026001}.

\bibitem{Kettemann04}
\Name{Kettemann S.} \Book{Distribution of the {K}ondo temperature in mesoscopic
  disordered metals} in \Book{Quantum Information and Decoherence in
  Nanosystems}, edited by \Name{Glattli D.~C., Sanquer M. \and Van J. T.~T.}
  (The Gioi Publishers) 2004 p. 259 (cond-mat/0409317).

\bibitem{Kettemann06}
\Name{Kettemann S. \and Mucciolo E.~R.} \REVIEW{Pis'ma v ZhETF }{83}{2006}{284}
  [JETP Letters {\bf 83}, 240 (2006)].

\bibitem{Kettemann07}
\Name{Kettemann S. \and Mucciolo E.~R.} \REVIEW{Phys. Rev. B
  }{75}{2007}{184407}.

\bibitem{Zhuravlev08}
\Name{Zhuravlev A., Zharekeshev I., Gorelov E., Lichtenstein A.~I., Mucciolo
  E.~R. \and Kettemann S.} \REVIEW{Phys. Rev. Lett. }{99}{2007}{247202}.

\bibitem{Affleck01}
\Name{Affleck I. \and Simon P.} \REVIEW{Phys. Rev. Lett. }{86}{2001}{2854}.

\bibitem{Simon03}
\Name{Simon P. \and Affleck I.} \REVIEW{Phys. Rev. B }{68}{2003}{115304}.

\bibitem{Simon06}
\Name{Simon P., Salomez J. \and Feinberg D.} \REVIEW{Phys. Rev. B
  }{73}{2006}{205325}.

\bibitem{KaulPRL06}
\Name{Kaul R., Zar\'and G., Chandrasekharan S., Ullmo D. \and Baranger H.}
  \REVIEW{Phys. Rev. Lett. }{96}{2006}{176802}.

\bibitem{Kaul09}
\Name{Kaul R.~K., Ullmo D., Zar\'and G., Chandrasekharan S. \and Baranger
  H.~U.} \REVIEW{Phys. Rev. B }{80}{2009}{035318}.

\bibitem{Pereira08}
\Name{Pereira R.~G., Laflorencie N., Affleck I. \and Halperin B.~I.}
  \REVIEW{Phys. Rev. B }{77}{2008}{125327}.

\bibitem{Kouwenhoven97}
\Name{Kouwenhoven L.~P., Marcus C.~M., McEuen P.~L., Tarucha S., Wetervelt
  R.~M. \and Wingreen N.~S.} \Book{Electron transport in quantum dots} in
  \Book{Mesoscopic Electron Transport}, edited by \Name{Sohn L.~L., Sch{\"o}n
  G. \and Kouwenhoven L.~P.} (Kluwer, Dordrecht) 1997 pp. 105--214.

\bibitem{BirdQdotsBook03}
\Name{Bird J.~P.} (Editor) \Book{Electron Transport in Quantum Dots} (Kluwer
  Academic Publishers, Dordrecht) 2003.

\bibitem{Lacroix79}
\Name{Lacroix C. \and Cyrot M.} \REVIEW{Phys. Rev. B }{20}{1979}{1969}.

\bibitem{Coleman83}
\Name{Coleman P.} \REVIEW{Phys. Rev. B }{28}{1983}{5255}.

\bibitem{Read84}
\Name{Read N., Newns D.~M. \and Doniach S.} \REVIEW{Phys. Rev. B
  }{30}{1984}{3841}.

\bibitem{BurdinRev2009}
\Name{Burdin S.} in \Book{Properties and Applications of Thermoelectric
  Materials}, edited by \Name{Zlatic V. \and Hewson A.} (NATO Science for Peace
  and Security Series B (9)) 2009 p. 325 (arXiv:0903.1942).

\bibitem{Bedrich10}
\Name{Bedrich R., Burdin S. \and Hentschel M.} \REVIEW{Phys. Rev. B
  }{81}{2010}{174406}.

\bibitem{Bohigas91}
\Name{Bohigas O.} \Book{Random matrix theories and chaotics dynamics} in
  \Book{Chaos and Quantum Physics}, edited by \Name{Giannoni M.~J., Voros A.
  \and Jinn-Justin J.} (North-Holland, Amsterdam) 1991 pp. 87--199.

\bibitem{MehtaBook}
\Name{Mehta M.~L.} \Book{Random Matrices (Second Edition)} (Academic Press,
  London) 1991.

\bibitem{us-future}
\Name{Liu D.~E., Burdin S., Baranger H.~U. \and Ullmo D.} in preparation
  (2011).

\end{thebibliography}
\end{document}